# Spin-to-Orbital Angular Momentum Conversion
# via Light Intensity Gradient

**Shuang-Yin Huang[1], Guan-Lin Zhang[1], Qiang Wang[1], Min Wang[1], Chenghou Tu[1], Yongnan Li[1,*], Hui-Tian Wang[2,3,*]**

**Abstract:** Besides a linear momentum, optical fields carry an angular momentum (AM), which have two intrinsic components: one is spin angular momentum (SAM) related to the polarization state of the field, and the other is orbital angular momentum (OAM) caused by the helical phase due to the existence of topological azimuthal charge. The two AM components of the optical field may not be independent of each other, and the Spin-to-Orbital AM conversion (STOC) under focusing will create a spin-dependent optical vortex in the longitudinal filed. Here we demonstrate a new mechanism (or novel way, new way, specific process) for the STOC based on a radial intensity gradient. The radial phase provides an effective way to control the local AM density, which induce counterintuitive orbital motion of isotropic particles in optical tweezers without intrinsic OAM. Our work not only provides fundamental insights into the spin-orbit interaction of light, but also push towards possible applications in optical micro-manipulation.

[1]Key Laboratory of Weak-Light Nonlinear Photonics and School of Physics, Nankai University, Tianjin 300071, China. [2]National Laboratory of Solid State Microstructures and School of Physics, Nanjing University, Nanjing 210093, China. [3]Collaborative Innovation Center of Advanced Microstructures, Nanjing University, Nanjing 210093, China. *e-mail: liyongnan@nankai.edu.cn; htwang@nju.edu.cn

Angular momentum (AM) is one of important characteristics of light[1,2] and has attracted increasing attention in a variety of applications, such as optical manipulations[3,4], quantum information[5-9], optical communications[10,11] and imaging[12-14]. For paraxial fields in free space, AM can be formally separated into a spin angular momentum (SAM) associated with the right- or left-hand circular polarization corresponding to positive or negative helicity of $\sigma = +1$ or $\sigma = -1$ and an orbital angular momentum (OAM) caused by a helical phase of $\exp(jm\phi)$, where $m$ is the azimuthal topological charge and $\phi$ is the azimuthal angle. In general, SAM and OAM can be experimentally distinguished according to their different mechanical actions on microparticles, because SAM usually makes the particles rotate around its own axis, while OAM causes orbital motion of the particles around the beam axis. This phenomenon is a conventional characterization method and distinguishing rule for SAM and OAM in experiment. However, in extreme cases, they cannot be separated. It has been known that Spin-to-orbit AM conversion (STOC) could occur when a circularly polarized Gaussian beam is converged by a lens[16,17], generating a spin-dependent optical vortex in the longitudinal field component. In fact, this conversion process verified in experiment depends mainly on the mechanical property of spin-dependent local AM (ref. 17), because the microparticles interact with the part of optical field.

For the circularly polarized optical field carrying the helical phase, the AM density in the direction of propagation is given by[15]

$$j_z = \varepsilon_0 \left[ \omega m \, | \, u \, |^2 - \tfrac{1}{2} \omega \sigma r \partial \, | \, u \, |^2 / \partial r \right], \tag{1}$$

where $u$ is a complex scalar function describing the distribution of the field amplitude, which satisfies the wave equation under the paraxial approximation. Obviously, in equation (1), the first term originates from the contribution of intrinsic OAM caused by the topological defect introduced by the helical phase in the field; while the second term arises from the combination of SAM ($\sigma$) and the radial intensity gradient (RIG). We usually change the azimuthal topological charge $m$ to effectively control the AM carried by the beam. In fact, changing the RIG may also provide an effective way to manipulate the sign and value of the local AM density and it may help us to deeply understand the STOC. But an analysis of the relation between the local AM density, the STOC and the RIG is still missing and remains a problem[16].

Here we provide an effective way to manipulate the RIG and harness local AM density. We predict in theory and validate in experiment a spin-dependent local AM associated with the RIG. Such a spin-dependent local AM can induce a counterintuitive orbital motion of isotropic particles in optical tweezers and indicate there is a new STOC mechanism. This breaks the cognition that STOC will create a spin-dependent optical vortex, in which the RIG plays an important role in the process.

## THEORETICAL ANALYSIS

To detect or characterize the AM experimentally, the optical trap or tweezers provides an effective method by the aid of exchange of mechanical torque acting on microparticles based on the transfer of AM. In previous works, the intensity gradient had been ignored because its symmetric distribution near the focus. For the circularly polarized Gaussian beam, the microparticle is usually trapped in the beam center results zero orbital rotation torque because the opposite RIG across the center. Even for the circularly polarized Laguerre-Gaussian beam, the probed dielectric microparticles will be stably trapped at the strongest ring with a defined radius, but the azimuthal force from the near odd-symmetric RIG is also zero. This implies the fact that the local AM density caused by the non-symmetric RIG will have contribution to the orbital motion of the spherical microparticle trapped at the strongest ring, but a key problem is how to achieve the non-symmetric RIG (or the non-odd-symmetric RIG) about the strongest ring.

The radial phase (RP) may provide an effective way to control the RIG. The RP can be classified into two categories: Category-(i), which contains the odd power of radial coordinate only, and Category-(ii), which contains the even power of radial coordinate only. Category-(i) is completely different from Category-(ii) that the odd (even) power of radial coordinate will (will not) dramatically change the spatial structure of the optical field. When the RP contains the even power of radial coordinate only[18,19], a spin-3/2 light has been predicted[18]. Here we devote to harness the local AM density based on the RIG, by using the linearly-varying RP without the intrinsic OAM caused by the topological defect of helical phase, unlike Ref. 17 requires the *tightly focused* circularly polarized *vortex*. Since the introduction of the linearly-varying (odd power) RP results in the substantial change of spatial structure of the optical field, for instance, a fundamental Gaussian beam will become a doughnut-like pattern.

Under the paraxial approximation, the transversal electric field of light with linearly-varying RP without the helical phase in free space can be written as follows

$$\mathbf{E}_\perp(r) \propto A(r)\exp\left( j2q\pi r\,/\,r_0 \right)[\hat{\mathbf{e}}_x + j\sigma\hat{\mathbf{e}}_y], \tag{2}$$

where $A(r)$ represents the amplitude of the optical field, $q$ is called the radial index that can be an any number in the range $[-\infty, +\infty]$ and can describe the radial gradient of phase, $r_0$ is the radius of the field, $\sigma$ represents the polarization state of light ($\sigma = \pm 1$ for right- and left-circularly polarized light and $\sigma = 0$ for linearly polarized). The linearly-varying RP $\exp(j2q\pi r/r_0)$ can be easily constructed by a spatial light modulator (SLM) acts as an equivalent "axicon", as shown in Figs. 1(a) and 1(b). If there is no "axicon" ($q = 0$), the input optical field will be focused in the Fourier plane (geometric focal plane) of the lens [$z = 0$, shown by the black dash-dot line in Figs. 1(a) and 1(b)]. In the presence of a linearly-varying RP, however, the true focal fields will be moved toward the $+z$ (for $q > 0$) and $-z$ (for $q < 0$) directions away from the Fourier plane of the lens, as shown in Figs. 1(a) and 1(b), respectively. For more details, we should numerically simulate the transverse distribution of focused fields using the Richards-Wolf integral[20,21] (see Supplementary Information). As shown in Figs. 1(c) and 1(d), the simulated $x$-$z$ plane intensity distributions in the vicinity of the Fourier plane of the lens with a focal length of $f = 500$ mm indeed confirm the fact analyzed in Figs. 1(a) and 1(b). We also simulate the transverse intensity distributions of the focused field at the three different planes [①, ② and ③ in Figs. 1(c) and 1(d)], which for both $q > 0$ and $q < 0$ exhibit the doughnut-like patterns similar to the focused vortex fields, as shown in Figs. 1(c1)-(c3) and 1(d1)-(d3). We define the radius of the strongest ring of the doughnut-like pattern as $R_P$. In the Fourier plane of the lens, the doughnut-like patterns for both $q > 0$ and $q < 0$ have the same intensity distribution [Figs. 1(c2) and 1(d2)], in particular, the transverse intensity distribution across the strongest ring with the radius of $R_P$ is symmetric (i.e., the RIG is odd symmetric). In the $z \neq 0$ plane, however, the transverse intensity distribution across the strongest ring is asymmetric. For instance, in the situation of $q > 0$, the doughnut-like pattern has the larger RIG within $r < R_P$ for $z < 0$ [Fig. 1(c1)] than within $r > R_P$ for $z > 0$ [Fig. 1(c3)]. The situation of $q < 0$ [Fig. 1(d1) and 1(d3)] is opposite to that of $q > 0$. This is completely different from the Laguerre-Gaussian field, which always exhibits a near symmetric intensity distribution with respect to the strongest ring, in any plane along the propagation direction. Clearly, the linearly-varying RP provides an effective method to control the RIG.

If the size of microparticles is comparable with the width of the ring of the focused optical field with linearly-varying RP, we consider the integral of the AM density $j_z$ across the ring (from $R_P - a$ to $R_P + a$) to describe the local interaction between the focused field and microparticles

$$J_z^L \propto \int_{R_P-a}^{R_P+a} j_z dr = \int_{R_P-a}^{R_P+a} \omega r \sigma \frac{\partial |A(r)|^2}{\partial r} dr = \omega \sigma \left[ \int_{R_P-a}^{R_P+a} r |A(r)|^2 dr - \frac{1}{2} r^2 |A(r)|^2 \Big|_{R_P-a}^{R_P+a} \right]. \qquad (3)$$

The gradient force constrains the particles to the strongest ring and $a$ is the radius of the probing particle. As shown in Fig. 1, in the Fourier plane of the lens, the local doughnut-like pattern has a Gaussian-like intensity distribution as $\exp[(r-R_P)^2/a^2]$, so the net local AM $J_z^L$ will be near zero, which cannot result in the orbital motion of microparticles. In the plane $z \neq 0$, however, $J_z^L$ will be nonzero due to the asymmetric radial intensity (i.e., non-odd symmetric RIG) about $r = R_P$, meaning that the trapped microparticles will move along a circular orbit. We have simulated the radial-varying intensity and AM density in the vicinity of the Fourier plane of the lens for the right-circularly polarized ($\sigma = +1$) optical fields with $q = \pm 10$, as show in Figs. 1(c1)-(c3) and 1(d1)-(d3). Simultaneously, if we choose the diameter of microparticles to be 30 μm, the simulated local integral AM values are $J_z^L = +0.57$ for $q = +10$ and $J_z^L = -0.54$ for $q = -10$, in the plane of $z = +15$ mm from the Fourier plane, respectively. Obviously, the microparticles will move along the circular orbit although the input optical field has only SAM without the intrinsic OAM caused by the helical phase. The sense of the orbital motion depends on the sign of the radial index $q$ and the chirality $\sigma$ of SAM. For given $q$ and $\sigma$, $J_z^L$ depends also on the longitudinal coordinate $z$, especially its sign will change across the Fourier plane of the lens, which is completely different from the intrinsic OAM causes by the helical phase. Of course, $J_z^L$ depends also on SAM. When the optical field is linearly polarized (i.e., $\sigma = 0$), the microparticles have no orbital motion no matter what intensity distribution of the focused field. For the nonzero local AM $J_z^L$ and then for the orbital motion of microparticles, therefore, SAM and the linearly-varying RP are both indispensable.

We can understand the above physical process by the simple phenomenological model of the spin flow schematically illustrated in Fig. 2. The physical ground is the energy circulation in the circularly polarized light[22]. We can imagine that the rotation of the field vectors takes place in "every point" of the focused field and the energy circulates within microscopic "cells". For the circularly polarized light with uniform intensity, all the "cells" are identical, the contributions of spin flow from the adjacent cells counterbalance and the macroscopic energy flow is absent. The compensation is not complete if the spin flow (shown by blue circles with arrows in Fig. 2, the larger and thicker blue circles with arrows indicate the stronger spin flow) of the adjacent cells are different (i.e. the intensity is transversely inhomogeneous), and the net spin flow between adjacent cells (red arrows in Fig. 2, the thicker red arrows indicate the stronger net spin flow) is along the azimuth, i.e. orthogonal to the radial intensity gradient. In the $z = 0$ plane, the doughnut-like field exhibits the Gaussian-like symmetric intensity distribution $\exp[(r-R_P)^2/a^2]$ about the strongest ring with the radius of $R_P$, as shown by the symmetric dark green doughnut-like in Fig. 2(a); this is, the radial intensity gradient is odd symmetric about $r = R_P$, so the spin flow is also odd symmetric about $r = R_P$. Therefore, the macroscopic spin flow vanishes, this is the reason why a trapped calcite particle only spins about own axis while has no orbital motion, which is similar to the circular polarized Laguerre-Gaussian beam[23]. In contrast, in the $z \neq 0$ plane, the optical field becomes asymmetric in intensity about $r = R_P$, as shown by the asymmetric dark green doughnut-like in Fig. 2(b); this is, the radial intensity gradient lacks the odd symmetry about $r = R_P$, so the spin flow has also no odd symmetry about $r = R_P$. As a result, the macroscopic spin flow becomes nonzero along the azimuth, as shown by open red arrow in Fig. 2(b), this is the reason why the trapped particles have orbital motion.

**EXPERIMENTAL RESULTS**

To confirm the above theoretical analysis and simulation results, we experimentally generate the optical fields with linearly-varying RP, using a spatial light modulator (SLM) loaded a hologram of blazed grating [see Method for details]. We measure the spatial intensity patterns of focused fields of the right-circular polarized optical fields with radial indices of $q = \pm 10$ in the vicinity of the Fourier plane of lens ($f = 500$ mm) by using a CCD, as shown by the insets in Figs. 3(a)-(f). For clarity, the blue solid curves in Figs. 3(a)-(f) plot the radial intensity distributions read from the images recorded by CCD. Correspondingly, the red dot curves in Figs. 3(a)-(f) plot the local AM densities calculated from the radial intensity distributions shown by blue solid curves in Figs. 3(a)-(f). Obviously, the experimental results in Fig. 3 are in good agreement with the simulated doughnut-like patterns in Fig. 1.

Optical tweezers, as an useful tool, has been successfully used to probe the AM flow by means of the interaction of microparticles with light[24,25]. Here we also perform the experiment of optical tweezers to validate the local AM property of the optical fields with linearly-varying RP (see Method for details). We are interested in realizing the counterintuitive orbital motion of the particles, therefore we choose neutral isotropic colloidal microspheres with a diameter of 3.2 μm as probed particles in experiment. To observe the obvious orbital motion of the trapped particles, we use an optical tweezers system composed of an objective (with a high numerical aperture of NA = 0.75) by an incident optical field with the radial index of $q = +6$ and a power of 108 mW. We also simulate the intensity distribution and AM density of its focused light (see Figs. S1-S4), which are very similar to the weak focusing situations mentioned above (Figs. 1 and 3). The close-packed nine particles are trapped in the doughnut-like behind the geometric focal plane ($z > 0$) of the objective, and exhibit clockwise or counterclockwise orbital motion, depending on the chirality of circular polarization, as shown in Figs. 4(a) and 4(b). For the linearly polarized light ($\sigma = 0$), no orbital motion of particles is observed. The maximum number of particles trapped in the doughnut-like can be controlled by changing the radial index $q$. For the case of $q = +7$ in Fig. 4(c), the trapped particles have increased to ten and move clockwise for the right-circularly polarized light; while when switching to $q = -7$ in Fig. 4(d), the motion direction of trapped particles is synchronously reversed. Obviously, the orbital motion of the trapped particles arises from the transfer of the local AM $J_z^L$ to particles through the RIG, which can be effectively controlled by the RP.

In addition, $J_z^L$ has also an important feature that is its longitudinal dependence. The trapped particles move clockwise for $q = +6$ and $\sigma = +1$ in the case of $z < 0$ in Fig. 5(a), while the sense of the orbital motion of the trapped particles is reversed for $q = +9$ and $\sigma = +1$ in the case of $z > 0$ in Fig. 5(b). More importantly, the local AM caused by the RIG can have the same or opposite sign as the intrinsic OAM originated from the vortex phase (see Fig. S5 for details). We can load the hologram with the linearly-varying RP of $q = +6$ and the vortex phase of $m = +1$ on SLM simultaneously, then the focused field exhibits still the doughnut shape. When $\sigma = +1$, the trapped particles almost stop behind the geometric focal plane ($z > 0$), the local AM from the linearly-varying RP of is compensated with the intrinsic OAM of helical phase, which means that the net total local AM acting on particles is almost zero in Fig. 6(a). Switching the polarization from $\sigma = +1$ to $\sigma = -1$, the trapped particles move clockwise with a shorter orbital period of ~8 s in Fig. 6(b).

**Discussion and Conclusion**

We experimentally demonstrate the fact that in the absence of intrinsic OAM, the net nonzero local AM density may not only be different in magnitude but also has the different sign from the SAM with the linearly-varying RP. The sense and velocity of orbital motion of microparticles depend on SAM and the RIG and this local AM can be continuously changed by selecting arbitrary radial index $q$. The effective torque on particles from the local $J_z^L$ is equivalent to that from the intrinsic OAM, and in this sense the radial phase provides a new way for optical manipulation. Our scheme with continuously control of the local AM density based on the linearly-varying RP is also compatible with other degree of freedom of light. The observed orbital motion of the isotropic particles in optical tweezers using the optical fields without intrinsic OAM may show the same the experiment phenomena as the OAM caused by the spin-to-orbital conversion from a tightly focused circularly polarized Gaussian beam[17] and an inhomogeneous and anisotropic metamaterial[26]. By numerical simulation, the intensity of the longitudinal field is very small comparing with transverse fields (see Fig. S6) and it indicates the RIG plays an important role in the STOC.

In conclusion, we have theoretically predicted and verified the STOC associated with the RIG through optical trapping experiments. The result breaks the limitation that orbital motion must be associated with the azimuthal phase gradient and enriches the category of AM. This spin-dependent AM will also give us a deeper understanding of SAM in optics as well as conversion of spin to orbit. Since this AM can be easily generated and arbitrarily tunable, it opens a novel route to control the light-matter interaction and optical metrology and it may show great applications in optical micro-manipulation and micro-fabrication. Besides, it is also expected to be found in other natural waves such as electron beams and acoustic waves[27-29].

**REFERENCES**

1. S. Franke-Arnold, L. Allen, and M. Padgett, ``Advances in optical angular momentum,'' Laser & Photonics Reviews **2**, 299 (2008).

2. A. M. Yao and M. J. Padgett, ``Orbital angular momentum: origins, behavior and applications,'' Advances in Optics and Photonics **3**, 161 (2011).

3. N. B. Simpson, K. Dholakia, L. Allen, and M. J. Padgett, ``Mechanical equivalence of spin and orbital angular momentum of light: an optical spanner,'' Opt. Lett. **22**, 52 (1997).

4. D. Gao, W. Ding, M. Nieto-Vesperinas, X. Ding, M. Rahman, T. Zhang, C. T. Lim, C. W. Qiu, ``Optical manipulation from the microscale to the nanoscale: fundamentals, advances and prospects,'' Light: Science & Applications **6**, 17039 (2017).

5. M. Erhard, R. Fickler, M. Krenn, A. Zeilinger, ``Twisted photons: new quantum perspectives in high dimensions," Light: Science & Applications **7**, 17146 (2018).

6. M. Malik, M. Erhard, M. Huber, M. Krenn, R. Fickler, A. Zeilinger, ``Multi-photon entanglement in high dimensions," Nature Photonics **10**, 248 (2016).

7. J. Yin, Y. H. Li, S. K. Liao, M. Yang, Y. Cao, L. Zhang, J. G. Ren, W. Q. Cai, W. Y. Liu, S. L. Li, R. Shu, Y. M. Huang, L. Deng, L. Li, Q. Zhang, N. L. Liu, Y. A. Chen, C. Y. Lu, X. B. Wang, F. Xu, J. Y. Wang, C. Z. Peng, A. K. Ekert, and J. W. Pan, ``Entanglement-based secure quantum cryptography over 1,120 kilometres," Nature **582**, 501 (2020).

8. J. G. Ren, P. Xu, H. L. Yong, L. Zhang, S. K. Liao, J. Yin, W. Y. Liu, W. Q. Cai, M. Yang, L. Li, K. X. Yang, X. Han, Y. Q. Yao, J. Li, H. Y. Wu, S. Wan, L. Liu, D. Q. Liu, Y. W. Kuang, Z. P. He, P. Shang, C. Guo, R. H. Zheng, K. Tian, Z. C. Zhu, N. L. Liu, C. Y. Lu, R. Shu, Y. A. Chen, C. Z. Peng, J. Y. Wang, and J. W. Pan, ``Ground-to-satellite quantum teleportation," Nature **549**, 70 (2017).

9. G. Vallone, V. D'Ambrosio, A. Sponselli, S. Slussarenko, L. Marrucci, F. Sciarrino, and P. Villoresi, ``Free-space quantum key distribution by rotation-invariant twisted photons," Phys. Rev. Lett. **113**, 060503 (2014).

10. J. Wang, ``Advances in communications using optical vortices," Photonics Research **4**, 14 (2016).

11. A. E. Willner, H. Huang, Y. Yan, Y. Ren, N. Ahmed, G. Xie, C. Bao, L. Li, Y. Cao, Z. Zhao, J. Wang, M. P. J. Lavery, M. Tur, S. Ramachandran, A. F. Molisch, N. Ashrafi, and S. Ashrafi, ``Optical communications using orbital angular momentum beams," Advances in Optics and Photonics **7**, 66 (2015).

12. L. Chen, J. Lei, and J. Romero, ``Quantum digital spiral imaging," Light: Science & Applications **3**, 153 (2014).

13. X. Fang, H. Ren, and M. Gu, ``Orbital angular momentum holography for high-security encryption," Nature Photonics **14**, 102 (2020).

14. H. Ren, X. Fang, J. Jang, J. Bürger, J. Rho, and S. A. Maier, ``Complex-amplitude metasurface-based orbital angular momentum holography in momentum space," Nature Nanotechnology **1**, 8 (2020).

15. L. Allen, M. W. Beijersbergen, R. J. C. Spreeuw, and J. P. Woerdman, ``Orbital angular momentum of light and the transformation of Laguerre-Gaussian laser modes," Phys. Rev. A **45**, 8185 (1992).

16. Y. Zhao, D. Shapiro, D. McGloin, D. T. Chiu, and S. Marchesini, ``Direct observation of the transfer of orbital angular momentum to metal particles from a focused circularly polarized Gaussian beam," Opt. Express **17**, 23316 (2009).

17. Y. Zhao, J. S. Edgar, G. D. M. Jeffries, D. McGloin, and D. T. Chiu, ``Spin-to-orbital angular momentum conversion in a strongly focused optical beam," Phys. Rev. Lett. **99**, 073901 (2007).

18. O. El Gawhary, T. Van Mechelen, and H. P. Urbach, ``Role of radial charges on the angular momentum of electromagnetic fields: spin-3/2 light," Phys. Rev. Lett. **121**, 123202 (2018).

19. Z. Man, Z. Xi, X. Yuan, R. E. Burge, and H. P. Urbach, ``Dual coaxial longitudinal polarization vortex structures," Phys. Rev. Lett. **124**, 103901 (2020).

20. B. Richards and E. Wolf, ``Electromagnetic diffraction in optical systems, II. Structure of the image field in an aplanatic system," Proceedings of the Royal Society of London. Series A. Mathematical and Physical Sciences **253**, 358 (1959).

21. X. Z. Gao, Y. Pan, G. L. Zhang, M. D. Zhao, Z. C. Ren, C. H. Tu, Y. N. Li, and H. T. Wang, ``Redistributing the energy flow of tightly focused ellipticity-variant vector optical fields," Photonics Research **5**, 640 (2017).

22. A. Bekshaev, K. Y. Bliokh, and M. Soskin, ``Internal flows and energy circulation in light beams," J. Opt. **13**, 053001 (2011).

23. A. T. O'neil, . MacVicar, L. Allen, and M. J. Padgett, ``Intrinsic and extrinsic nature of the orbital angular momentum of a light beam," Phys. Rev. Lett. **88**, 053601 (2002).

24. M. J. Padgett and R. Bowman, ``Tweezers with a twist," Nature photonics **5**, 343 (2011).

25. X. L. Wang, J. Chen, Y. Li, J. Ding, C. S. Guo, and H. T. Wang, ``Optical orbital angular momentum from the curl of polarization," Phys. Rev. Lett. **105**, 253602 (2010).

26. M. Kang, J. Chen, X. L. Wang, and H. T. Wang, ``Twisted vector field from an inhomogeneous and anisotropic metamaterial," J. Opt. Soc. Am. B **29**, 572-576 (2012).

27. S. M. Lloyd, M.Babiker, G.Thirunavukkarasu, and J. Yuan, ``Electron vortices: Beams with orbital angular momentum," Reviews of Modern Physics **89**, 035004 (2017).

28. A. Ozcelik, J. Rufo, F. Guo, Y. Gu, P. Li, J. Lata, and T. J. Huang, ``Acoustic tweezers for the life sciences," Nature methods **15**, 1021 (2018).

29. X. Jiang, Y. Li, B. Liang, J. C. Cheng, and L. Zhang, ``Convert acoustic resonances to orbital angular momentum,'' Phys. Rev. Lett. **117**, 034301 (2016).

**Acknowledgements**

**Funding:** This work was supported by the National key R&D Program of China (2017YFA0303800, 2017YFA0303700), the National Natural Science Foundation of China (12074197, 11774183, 11674184).

**Author contributions**

S.Y.H, C.H.T., Y.N.L. and H.T.W. designed the experiments; S.Y.H dominantly and G.L.Z., Q.W. and M.W. assistantly carried out the experiments and analyzed the data; Y.N.L. and H.T.W. planed and supervised the project. All authors contributed to the manuscript.

**Competing interests**

The authors declare no competing interests.

**Additional information**

Supplementary information is available for this paper at https://doi.org/xxxxxxxxx.

Reprints and permissions information is available at www.nature.com/reprints. Correspondence and requests for materials should be addressed to Y.N.L. or H.T.W. Publisher's note: Springer Nature remains neutral with regard to jurisdictional claims in published maps and institutional affiliations.

## METHODS

### Generation of the topological optical fields with linearly-varying radial phase

In our experiment, we create the topological optical fields with linearly-varying radial phase by using a phase-only spatial light modulator (SLM, a reflective phase modulator array of 1920×1080 pixels). The transmission function of the blazed grating on SLM is $t(x,y) = \gamma \operatorname{mod}(2\pi f_0 x + \delta, 2\pi)$, where $\delta$ is the additional phase distribution, $f_0$ and $\gamma$ are the spatial frequency and modulation depth of SLM, respectively (Fig. S7a and b). For the linearly polarized light, its first-order diffracted light has the phase $\exp(+j\delta)$ with a diffraction efficiency of ~70%. We set $\delta = \alpha_0 + 2\pi q r / r_0$ for the radial phase, where $r$ represents the radial coordinate and $r_0$ is the radius of the optical field incident on SLM. $\alpha_0$ and $q$ are the initial phase and the radial index, respectively. Interfering with the Gaussian beam, the intensity with concentric rings pattern experimentally verifies this property (Fig. S7c and d).

### Optical tweezers setup

Optical tweezers are capable of manipulating micrometer-sized dielectric particles. Our optical tweezers setup is based on a Zeiss inverted microscope as shown in Supplementary Information Fig S7. A linearly polarized continuous-wave laser (Verdi-5, Coherent Inc.) with a power of 5 W and a wavelength of 532 nm is expanded and collimated by a telescope and is then reflected onto the phase-only SLM (Pluto II, HOLOEYE Photonics AG, Germany). The required optical field is created by the computer generated hologram loaded on SLM, and then passes through a 4f system consisting of two lenses (L1 and L2 with the same focal length of $f = 250$ mm). The input plane of the objective O1 and the SLM plane are the conjugate image planes of the 4f system with each other. A filter at the Fourier plane of the 4f system blocks all diffraction orders except for the first order. A quarter-wave plate (QWP) is inserted between L2 and O1 to convert the linearly polarized light into the circularly polarized one. Optical trapping is performed using the inverted microscope (Zeiss observer Z5) with the objective O1 (60× and NA = 0.75). The particles used for optical trapping in the experiment are neutral isotropic colloidal beads of 3.2 μm in diameter and were dispersed in sodium dodecyl sulfate solution in the cell with glass coverslip. The image of the trapped particles are observed another objective O2 (100× and NA = 0.95). Two charge coupled device cameras CCD1 and CCD2 with resolution of 1280×1024 pixels and maximal frame rate of 60 fps are used to record the manipulation process of the trapped particles.

**Figure Legend**

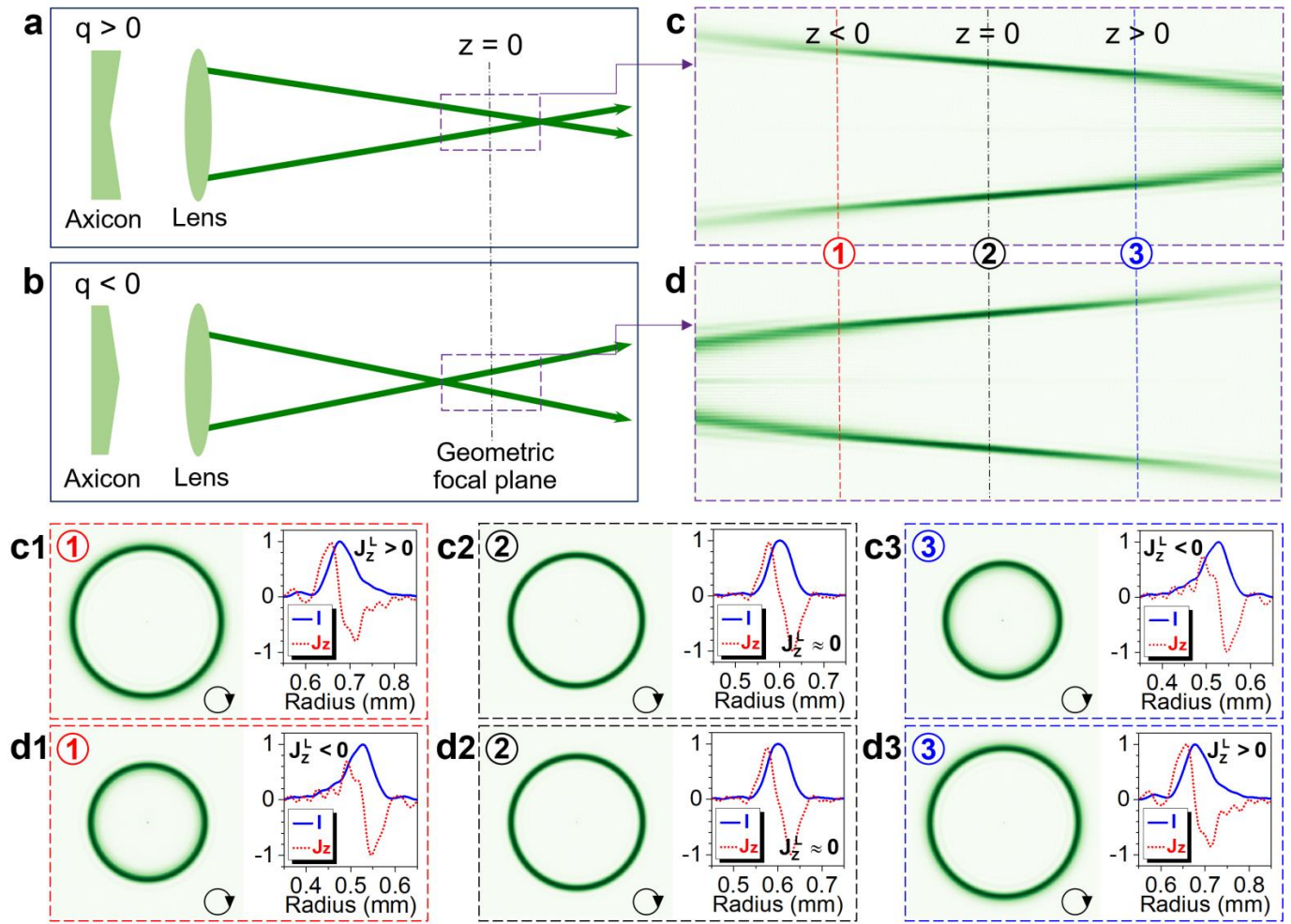

**Fig. 1** Simulated intensity distributions and local AM densities of focused optical fields with the linearly-varying radial phase for the right-hand circular polarization $\sigma = 1$. **a** (**b**), Schematic diagram of focusing process of optical fields with the radial index of $q > 0$ ($q < 0$). $z = 0$ represents the geometric focal plane (Fourier plane) of the lens. **c** (**d**), Simulated $x$-$z$ plane intensity distribution of focused optical field with the radial index $q = 10$ ($q = -10$) in the vicinity of the geometric focal plane of the lens with a focal length of $f = 500$ mm within the range of $z \in [-30, 30]$ mm. **c1-c3** (**d1-d3**), Simulated transverse intensity distributions and local AM densities at three different planes shown in **c** (**d**): ① $z = -15$ mm, ② $z = 0$, ③ $z = 15$ mm, in which all the pictures have the same size of $1.76 \times 1.76$ mm².

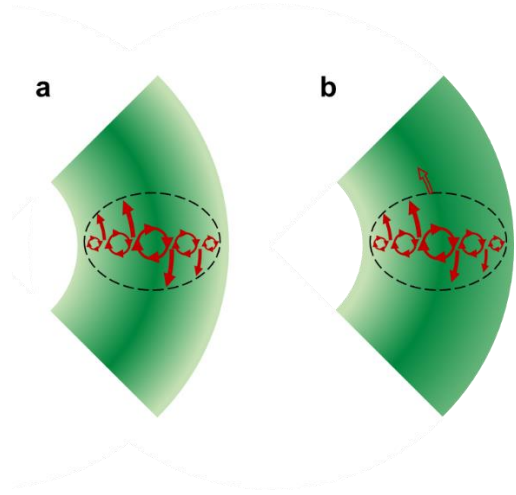

**Fig. 2** Schematic diagrams of emergence of the azimuthal force induced by the intensity gradient and circular polarization. The green color present the intensity distribution of the annular focus ring: **a** symmetric and **b** asymmetric. Red circles present the circulation cells of the spin flow at different position and their size is proportional to the local intensity. The red arrows present the net force by uncompleted compensation of the adjacent circulation cells. The open red arrow presents the azimuthal resultant force if we consider the particle as a rigid body.

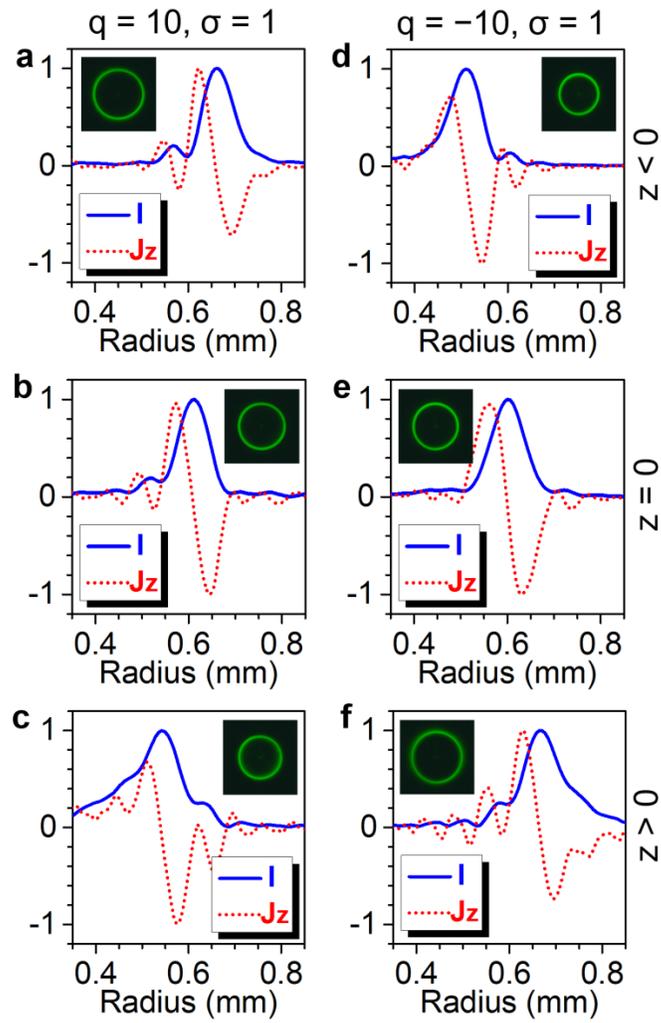

**Fig. 3** Experimentally measured intensity distribution (blue curve) and corresponding calculated AM density (red dots) around the Fourier plane of lens ($f$ = 500 mm) with effective NA=0.01. **a-c** Radial index $n$ = 10. **d-f** Radial index n = -10. The polarization is right circular σ=1. Insets: the intensity distributions.

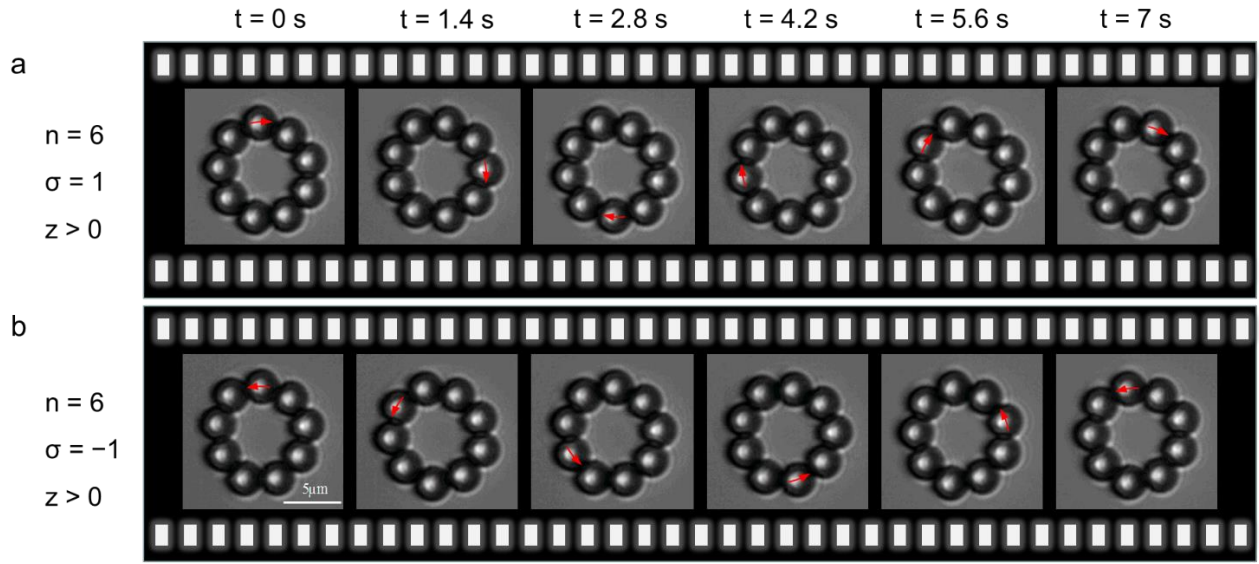

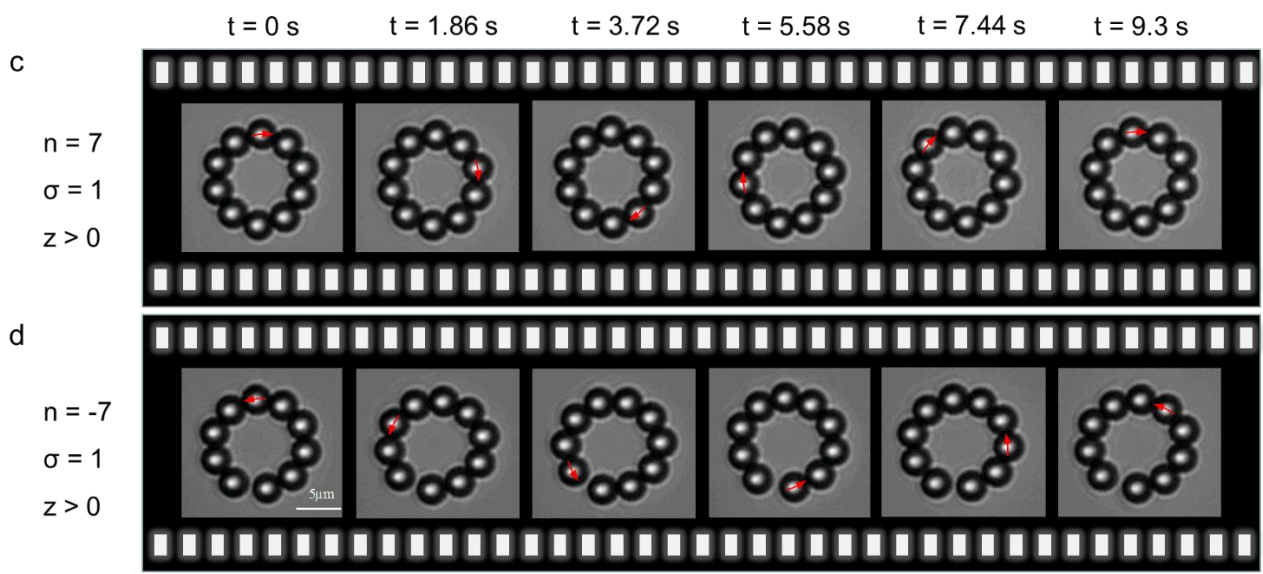

**Fig. 4** Camera snapshots of continuously rotating microparticles trapped by the focused light with the linearly-varying radial phase after the focal plane (z > 0) of the objective (NA = 0.75). **a**, $n = 6$ and $\sigma = 1$; **b**, $n = 6$ and $\sigma = -1$; **c**, $n = 7$ and $\sigma = 1$; **d**, $n = -7$ and $\sigma = 1$. The scale bar represents 5μm, and the red arrow denotes the orbital rotational direction. The corresponding elapsing time $t$ are denoted in each snapshot, and there is a video that can be seen in movies 1 and 2.

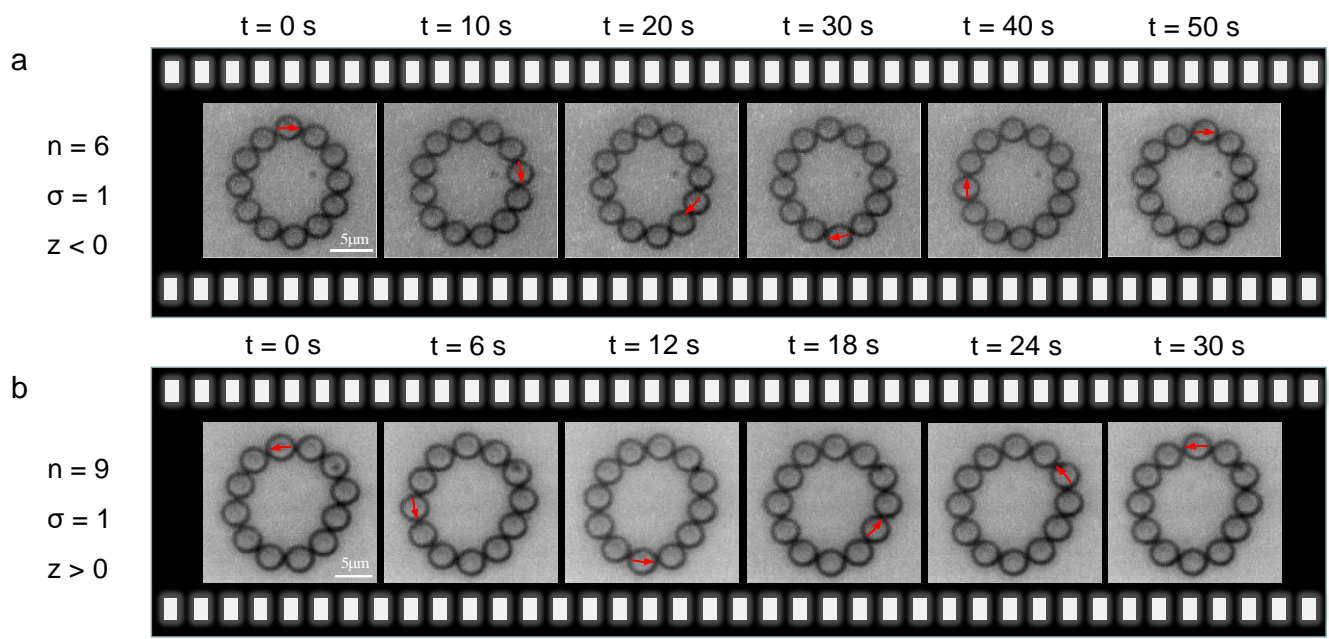

**Fig. 5** Camera snapshots of continuously rotating microparticles trapped by the focused light with the linearly-varying radial phase and circular polarization σ=1 for different position. **a**, before the focal plane (z < 0), $n$=6. **b**, behind the focal plane (z > 0), $n$=9. The NA of the objective is 0.75. The scale bar represents 5μm, and the red arrow denotes the orbital rotational direction. The corresponding elapsing time $t$ are denoted in each snapshot, and there is a video that can be seen in movie 4.

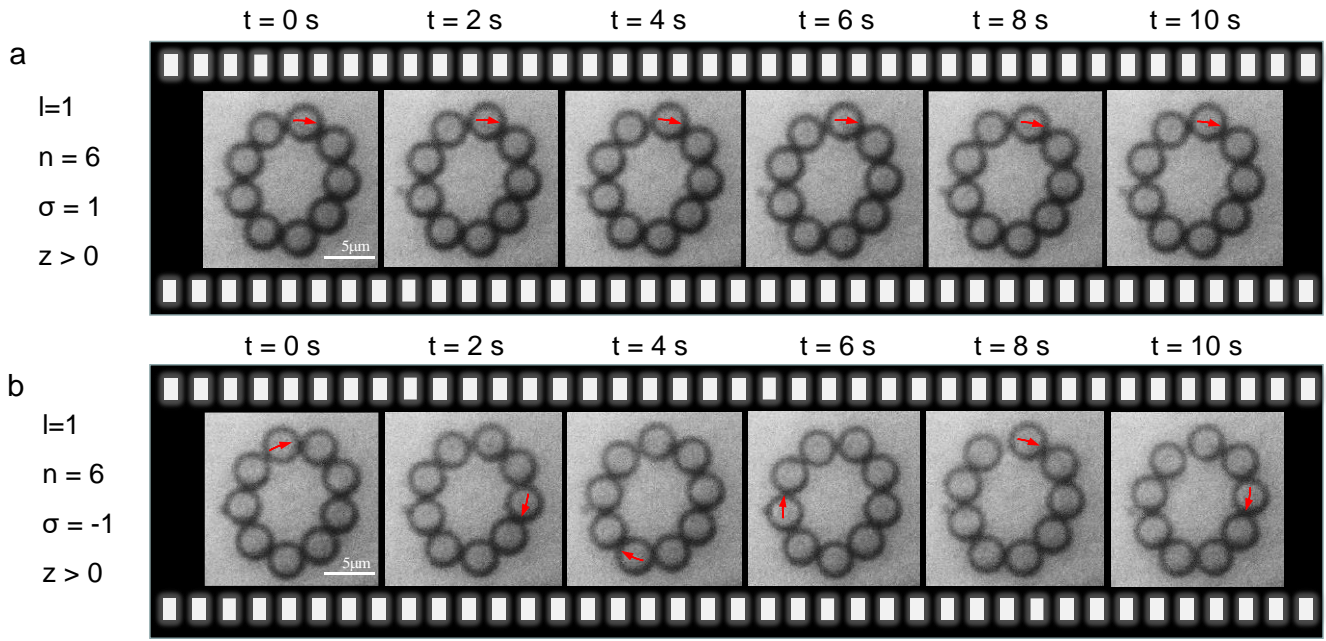

**Fig.6** Camera snapshots of continuously rotating microparticles trapped by the focused light with the azimuthal and radial phase after the focal plane (z > 0) of the objective (NA = 0.75) for different polarization. **a**, $\sigma = 1$ and **b**, $\sigma = -1$. The topological charge $l$=1, radial index $n$=6. The NA of the objective is 0.75. The scale bar represents 5μm, and the red arrow denotes the orbital rotational direction. The corresponding elapsing time $t$ are denoted in each snapshot, and there is a video that can be seen in movie 5.